\documentclass[aps,prl,twocolumn,nofootinbib]{revtex4-1}

\usepackage{graphicx}
\usepackage{dcolumn}
\usepackage{bm}
\usepackage{ textcomp }
\usepackage{color}
\usepackage{amsmath}
\usepackage{amssymb}
\usepackage{xcolor}
\usepackage{hyperref}
\usepackage{easyReview}
\usepackage{mathdots}
\usepackage{float}
\usepackage{lineno}
\usepackage{array}
\usepackage{xcolor,colortbl}
\usepackage[normalem]{ulem}

\definecolor{LightBlue}{rgb}{0.8,0.8,0.8}

\usepackage{amsmath} 

\allowdisplaybreaks

\DeclareUnicodeCharacter{2212}{-}
\begin{document}
\title{Static and dynamical magnetic properties of the extended Kitaev-Heisenberg model with spin vacancies}

\date{\today}

\author{Shaozhi Li}
\affiliation{Materials Science and Technology Division, Oak Ridge National Laboratory, Oak Ridge, TN 37831, USA}
\author{Randy S. Fishman}
\affiliation{Materials Science and Technology Division, Oak Ridge National Laboratory, Oak Ridge, TN 37831, USA}
\author{Tom Berlijn}
\affiliation{Center for Nanophase Materials Sciences, Oak Ridge National Laboratory, Oak Ridge, TN 37831, USA}

\begin{abstract}
Motivated by the potential to suppress the antiferromagnetic long-range order in favor of the long-sought-after Kitaev quantum spin liquid state, we study the effect of spin vacancies in the extended Kitaev-Heisenberg model. In particular, we focus on a realistic model obtained from fitting inelastic neutron scattering on $\alpha$-RuCl$_3$. We observe that the long-range zigzag magnetic ordered state only survives when the doping concentration is smaller than 5\%. Upon further increasing the spin vacancy concentration, the ground state becomes a short-range ordered state at low temperatures. Compared with experiments, our classical solution over-stabilizes the zigzag correlation in the presence of spin vacancies. Our theoretical results provide guidance toward interpreting inelastic neutron scattering experiments on magnetically diluted Kitaev candidate materials
\end{abstract}

\maketitle

\let\thefootnote\relax\footnotetext{
Copyright  notice: This  manuscript  has  been  authored  by  UT-Battelle, LLC under Contract No. DE-AC05-00OR22725 with the U.S.  Department  of  Energy. The  United  States  Government  retains  and  the  publisher,  by  accepting  the  article  for  publication, acknowledges  that  the  United  States  Government  retains  a  non-exclusive, paid-up, irrevocable, world-wide license to publish or reproduce the published form of this manuscript, or allow others to do so, for United States Government purposes.  The Department of Energy will provide public access to these results of federally sponsored  research  in  accordance  with  the  DOE  Public  Access  Plan (http://energy.gov/downloads/doe-public-access-plan)}

\section{Introduction}
The fractionalization of electrons induced by quantum many-body effects is one of the central topics in condensed matter physics. A celebrated representative example is the fractional quantum Hall effect in a two-dimensional electron system, where quasiparticles have a fractional electron charge~\cite{LaughlinPRL1983,ShengNature2011,CohenNature2019}. Another well-known example is the quantum spin liquid (QSL), where spins do not form an ordered state down to zero temperature and spin excitations are fractionalized into spinons and visons~\cite{ReadPRL1991,PunkNaturePhysics,SonnenscheinPRB2017}.
Over the past decades, a variety of QSL candidate materials have been found, including $\kappa$-(BEDT-TTF)$_2$Cu$_2$(CN)$_3$~\cite{ShimizuPRL,ShaozhiPRR2020}, EtMe$_3$Sb[Pd(dmit)$_2$]$_2$~\cite{Yamashita1246,KohlerPRB2008}, YbMgGaO$_4$~\cite{YueshengPRL,Shen2016nature,Paddison2017,shen2018naturecom,ShaozhiPRB2021}, and Sr$_2$CuO$_3$~\cite{FujisawaPRB1999,Schlappanature2012}.

The Kitaev model provides a good platform to study QSLs and fractional quasiparticles because its ground state can be obtained exactly~\cite{KITAEV20062}. To realize the exotic properties of the Kitaev model, there has been a keen interest in discovering Kitaev physics in real materials. The iridium oxides $A_2$IrO$_3$ ($A=$Li, Na)~\cite{ChaloupkaPRL2010,SinghPRB2010,SinghPRL2012,FoyevtsovaPRB2013,ChaloupkaPRL2013,YamajiPRL2014,WinterPRB2016} and the ruthenium compound $\alpha$-RuCl$_3$~\cite{PlumbPRB2014,KubotaPRB2015,SandilandsPRL2015,JohnsonPRB2015,SearsPRB2015,CaoPRB2016,KoitzschPRL2016,YamadaPRL2017,Kasahara2018nature,ShaozhiPRB2022} with strong spin-orbit coupling have been proposed as Kitaev candidate materials, where  fractional Majorana quasiparticles could be observed. More recently, Kitaev candidate materials have also been proposed based on honeycomb layers of Co$^{2+}$ or Ni$^{3+}$ ions~\cite{Motome_2020} and rare-earth chalcohalides~\cite{Ji_2021}.
However, due to stacking faults, the presence of Heisenberg, off-diagonal interactions, spin-phonon interactions~\cite{ShaozhiPRB2022}, and disorder, the ground states in these materials are distinct from the ground state of the Kitaev model. To understand the microscopic nature of these materials, it is necessary to carefully study complicated interactions that go beyond the Kitaev model.

The magnetic ground state of $\alpha$-RuCl$_3$ and numerous other Kitaev spin liquid candidate materials is the antiferromagnetic (AFM) zigzag state. One route to suppress this AFM state in favor of a potential quantum spin liquid state is the application of magnetic fields~\cite{YokoiScience2021}. Meanwhile, the AFM can also be suppressed by spin vacancies.
For example, $\alpha$-RuCl$_3$ exhibits zigzag magnetic order below 7 K, but incorporating Ir$^{3+}$ into Ru$_x$Ir$_{1-x}$Cl$_3$ reduces the N{\'e}el temperature of the ordered state~\cite{LampenPRL2017,SeungPRB2018,SeungPRL2020,BaekPRB2020}. Moreover, powder inelastic neutron scattering (INS) on Ru$_x$Ir$_{1-x}$Cl$_3$ shows that spin-vacancies leave intact spectral features associated with fractional excitations up to $x=0.35$~\cite{LampenPRL2017}. Other diluted Kitaev candidate materials being studied include: Ru$_x$Rh$_{1-x}$Cl$_3$~\cite{BastienPRM2022},  Na$_2$Ir$_{1-x}$Ti$_x$O$_3$, Li$_2$Ir$_{1-x}$Ti$_x$O$_3$~\cite{ManniPRB2014}, and Na$_2$Co$_{2−x}$Zn$x$TeO$_6$~\cite{FuPRB2023}.

Motivated by the potential of spin vacancies to suppress AFM in favor of the long-sought-after Kitaev quantum spin liquid,
our work studies their impact on the static and dynamic properties of the extended Kiteav-Heisenberg model. Previous relevant theoretical studies of this problem mostly focused on static properties of the pure Kitaev model~\cite{WillansPRB2011,NasuPRB2020,NasuPRB2021,KaoPRX2021} or the pure Kitaev-Heisenberg model~\cite{AndradePRB2014}. Our work studies the influence of spin vacancies on the static and dynamical magnetic properties of a previously published extended Kitaev-Heisenberg model obtained from fits against inelastic neutrons scattering on $\alpha$-RuCl$_3$~\cite{SamarakoonPRR2022}.
We first study the phase transition in the presence of spin vacancies using the replica exchange Monte Carlo method.  This work reveals that long-range zigzag order vanishes as the doping concentration reaches 5\%, and then the ground state exhibits short-range order. Tracing the change of the dynamical magnetic structure factor with different vacancy concentrations, we find that the low energy magnon mode persists up to a concentration that is larger than the site percolation threshold. Compared to experimental results, the zigzag correlation is over-stabilized in our classical solution.

\section{Model}
While various spin models have been proposed to describe the magnetic properties of $\alpha$-RuCl$_3$~\cite{Laurellnpj2020}, 
we focus on a model derived from fitting classical spin Hamiltonians against inelastic neutron scattering of $\alpha$-RuCl$_3$ via machine learning techniques~\cite{SamarakoonPRR2022}. This choice is more suitable, given the classical treatment of the spins in our work.

In this model, the spin-1/2 extended Kitaev-Heisenberg spin Hamiltonian on the honeycomb lattice is given by
\begin{eqnarray}
H&=&\sum_{\gamma=x,y,z}\sum_{\langle i,j \rangle_\gamma } {\bf S}_i
\cdot J_1^\gamma \cdot {\bf S}_j \nonumber\\
&+&J_2\sum_{\langle\langle i,j\rangle \rangle} {\bf S}_i\cdot {\bf S}_j + J_3\sum_{\langle \langle \langle i,j \rangle \rangle \rangle} {\bf S}_i\cdot {\bf S}_j,
\end{eqnarray}
where $\langle \cdots \rangle$, $\langle\langle \cdots \rangle\rangle$, and $\langle\langle\langle \cdots \rangle \rangle \rangle$ represent nearest, next-nearest, and third-nearest neighbors, respectively. The nearest neighbor exchange interaction matrix is defined as
\begin{eqnarray}
J_1^x&=&\begin{bmatrix}
J_1+K & 0 & 0 \\
0 & J_1 & \Gamma \\
0 & \Gamma & J_1
\end{bmatrix},\\
J_1^y&=&\begin{bmatrix}
J_1 & 0 & \Gamma \\
0 & J_1+K & 0 \\
\Gamma & 0 & J_1
\end{bmatrix},\\
J_1^z&=&\begin{bmatrix}
J_1 & \Gamma & 0 \\
\Gamma & J_1 & 0 \\
0 & 0 & J_1+K
\end{bmatrix}
\end{eqnarray}
These complicated interactions arise from the multiorbital nature of Ru $t_{2g}$ orbitals and the oxygen-atom mediated hopping~\cite{RauPRL2014,kumar2022}.
By fitting INS experimental data, it is found that $J_1=-0.4$ meV, $K=-5.3$ meV, $\Gamma=-0.15$ meV, $J_2=-0.19$ meV, and $J_3=1.35$ meV~\cite{SamarakoonPRR2022}.
In general, magnetic vacancies can change the electron transport around its neighboring sites. Since the exchange interaction is relevant to the hopping between two sites, magnetic vacancies can also impact the exchange interaction. However, determining this effect is complicated. In our work, we treat the most important influence of spin vacancies and neglect the modification of the exchange interactions between non-vacant sites.

\section{Replica exchange Monte Carlo method}
 We first study the spin model using the classical replica exchange Monte Carlo (MC) method with $2L\times L$ sites on the two-dimensional honeycomb lattice, which is spanned by the primitive vector ${\bf a}_1=a(\sqrt{3},0)$ and ${\bf a}_2=a(\frac{\sqrt{3}}{2},\frac{3}{2})$, where $a$ is the distance between two nearest neighbor Ru sites. The vacancy is simulated by randomly selecting a fraction $x$ of spins; consequently, the total number of spins is $N_s=(1-x)2L^2$. We performed swap updates between two different temperatures every 2 MC sweeps, and both the single-site heat bath and overrelaxation updates were used in each MC sweep. In our simulations, 50000 MC sweeps are used to thermalize the system, and $10^5$ MC sweeps are used to perform measurements. Disorder averages are taken over $N_r$ samples, with $N_r$ ranging from 100 for $x<0.1$ to $N_r=200$ for $x>0.1$.

For a physical system, the specific heat can accurately determine a second-order phase transition. In our simulations, the specific heat $C_p$ is computed via
\begin{eqnarray}
C_p=\frac{1}{N_s}\frac{1}{N_r}\sum_{r}\frac{\langle E^2 \rangle_{r,\text{MC}}-\langle E\rangle^2_{r,\text{MC}}}{k_B^2T^2},
\end{eqnarray}
where $\langle \cdots \rangle_{r,\text{MC}}$ is the average value over the MC samples for the $r$-th random vacancy configuration. $E$ represents the total energy.

In addition, we study the evolution of the magnetic correlation length to determine the temperature $T_N$ for the long-range ordered transition. Near the critical temperature, the spin correlation function $\chi_m({\bf q})$ in momentum space can be represented by
\begin{eqnarray}
\chi_m({\bf q})=\frac{C}{|{\bf q}-{\bf Q}_m|^2+\xi^{-2}},
\end{eqnarray}
where $\xi$ is the correlation length, and ${\bf Q}_m$ is the magnetic wave vector. On the finite lattice, we estimate the correlation length via
\begin{eqnarray}\label{Eq:correlation}
\xi^2=\frac{1}{4[\text{sin}^2(k_m^x/2)+\text{sin}^2(k_m^y/2)]}\left[\frac{\chi_m({\bf Q}_m)}{\chi_m({\bf Q}_m+{\bf k}_m)}-1 \right],
\end{eqnarray}
where ${\bf k}_m$ is the minimum allowed wave vector~\cite{BallesterosPRB2000}. In our calculations, we set ${\bf k}_m=(0,\frac{4\pi}{3La})$. In the presence of the disorder, $\chi_m({\bf q})$ in Eq.~\ref{Eq:correlation} is obtained by
\begin{eqnarray}
\chi_m(q)= \frac{1}{N_r}\sum_r \langle \chi_m(q) \rangle_{r,\text{MC}}.
\end{eqnarray}

In the thermodynamic limit, the correlation length diverges at the critical temperature as $|T-T_c|^{-v}$, where $v$ is the critical exponent and equals 1 for a two-dimensional Ising model. On a finite-size lattice, the correlation length is taken over by the lattice size, $\xi\sim cL$, where $c$ is a constant value.

\begin{figure}[t]
\center\includegraphics[width=\columnwidth]{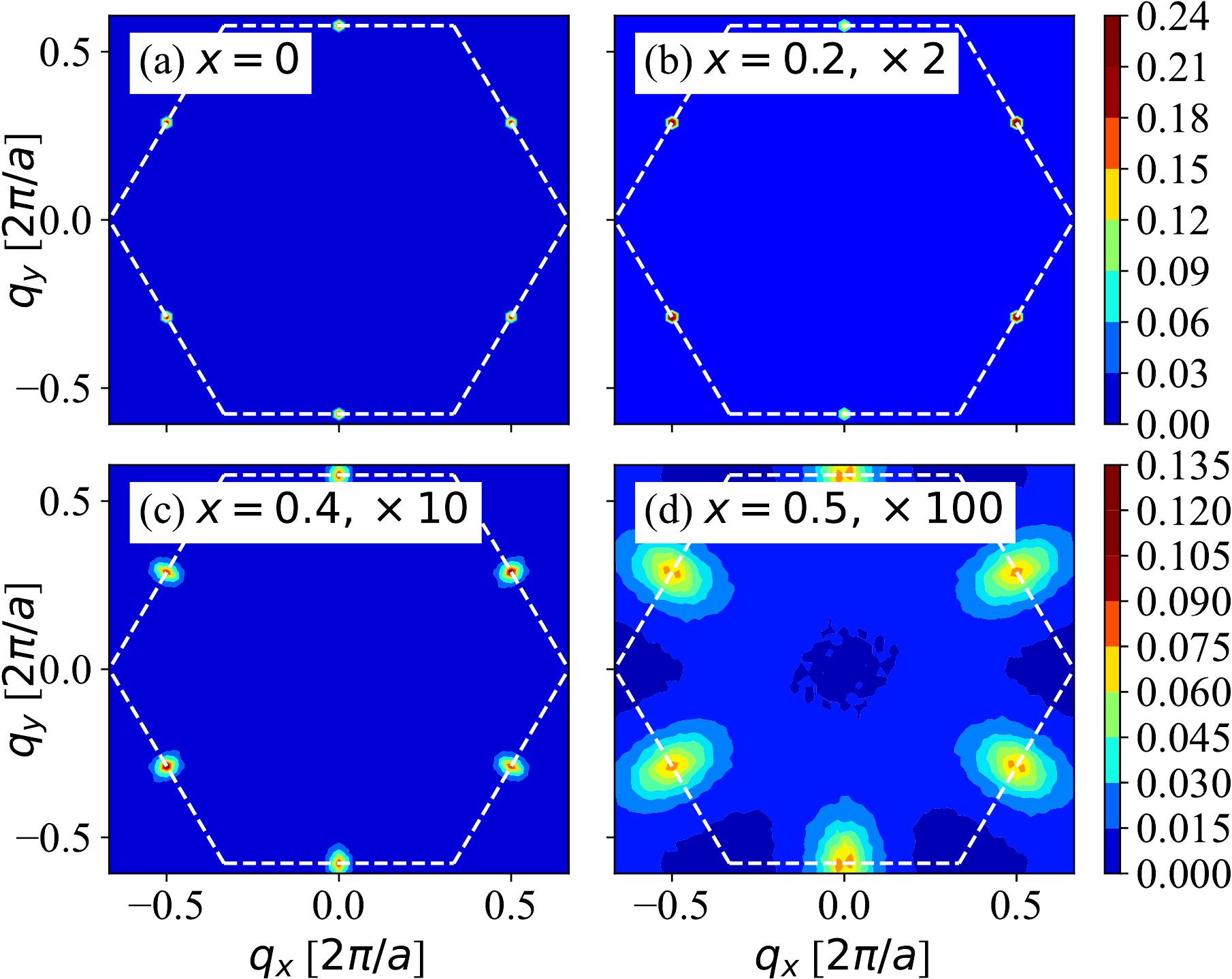}
\caption{\label{Fig:fig1} Static sublattice spin correlations $\chi_m(q)$ in the momentum space. Panels (a), (b), (c), and (d) plot the spin correlation function for $x=0$, 0.2, 0.4, and 0.5, respectively. $\chi_m(q)$ in panels (b), (c), and (d) is normalized by 2, 10, and 100, respectively. The white dashed line denotes the first Brillouin zone.}
\end{figure}

\begin{figure}[t]
\center\includegraphics[width=\columnwidth]{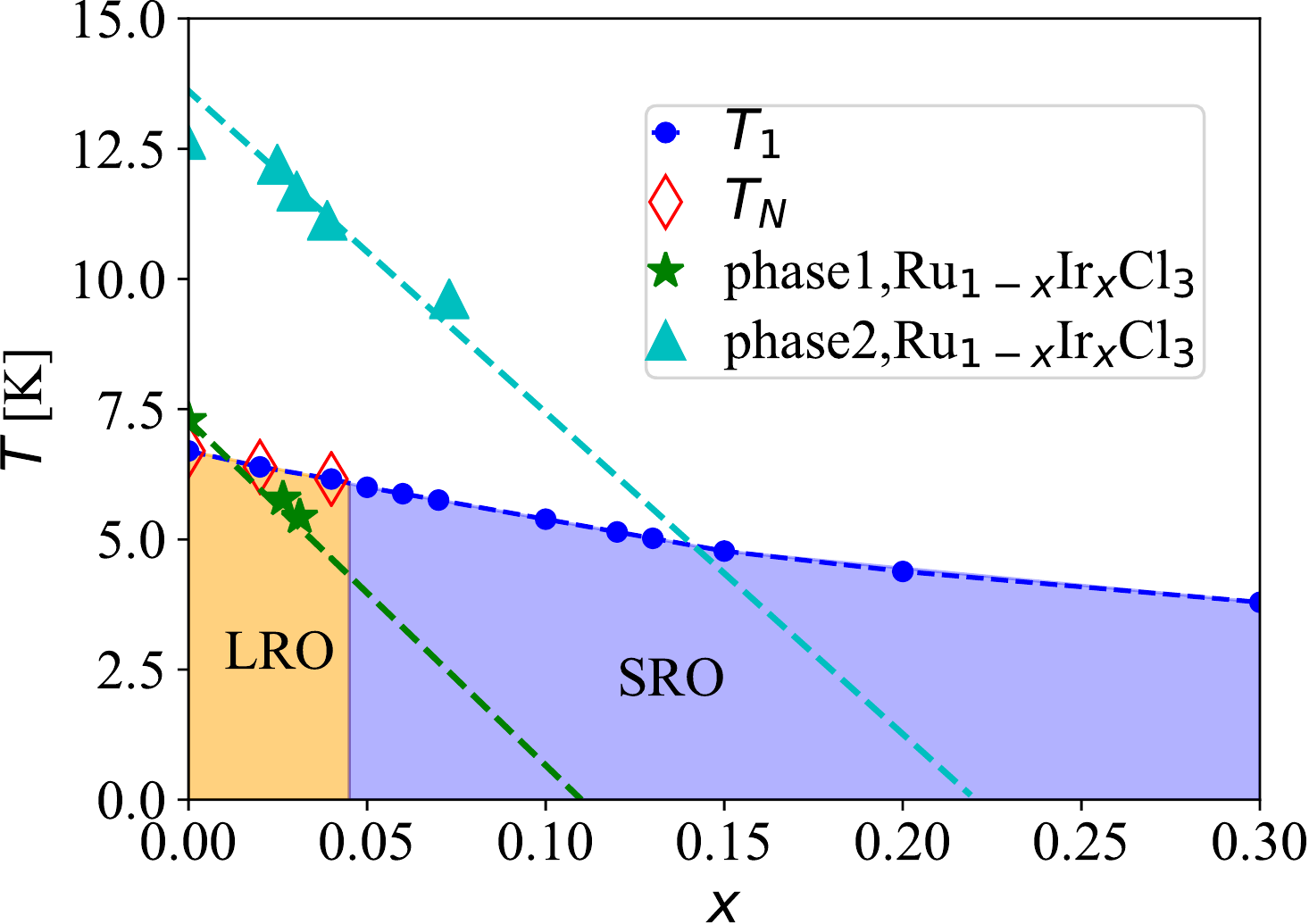}
\caption{\label{Fig:fig2} Phase diagram in the temperature $T$ and doping concentration $x$ plane. The blue symbol denotes the temperature $T_1$ where the specific heat has the maximum value. The red diamond denotes the ordering temperature from the scaling analysis. The cyan triangle and the blue star denote $T_1$ of two phase transitions in Ru$_{1-x}$Ir$_x$Cl$_3$, extracted from Ref.~\cite{LampenPRL2017}. LRO stands for the long-range order in the thermaldynamic limit. SRO stands for the short-range order.}
\end{figure}

\section{Langevin equation of motion}
To study the spin dynamics, we use the Langevin equation, which has the form
\begin{eqnarray}
\frac{d{\bf S}_i}{dt}=\frac{1}{\hbar}\left[{\bf S}_i \times ({\bf F}_i + {\bf f}_i) - \gamma {\bf S}_i \times \left({\bf S}_i \times {\bf F}_i \right)\right],
\end{eqnarray}
where ${\bf F}_i=-\partial H/\partial {\bf S}_i$ is the effective field acting on spin ${\bf S}_i$, and $H$ is the Hamiltonian. $\gamma$ is the dimensionless damping parameter. ${\bf f}_i(t)$ is a delta-correlated fluctuating effective magnetic field, satisfying the conditions $\langle {\bf f}_i(t)\rangle=0$ and $\langle f_{i,\alpha}(t)f_{j,\beta}(t^\prime)\rangle=\mu \delta_{ij}\delta_{\alpha\beta}\delta(t-t^\prime)$. Subscripts $\alpha$ and $\beta$ denote the Cartesian components of a vector. $\gamma$ and $\mu$ are related via $\mu=2\gamma \hbar k_B T$. The results shown in this work are computed with $\gamma=0.05$. We also run simulations with different values of $\gamma$ ($\gamma=0.01$ and 0.1) and find our current results are robust.

\begin{figure}[t]
\center\includegraphics[width=\columnwidth]{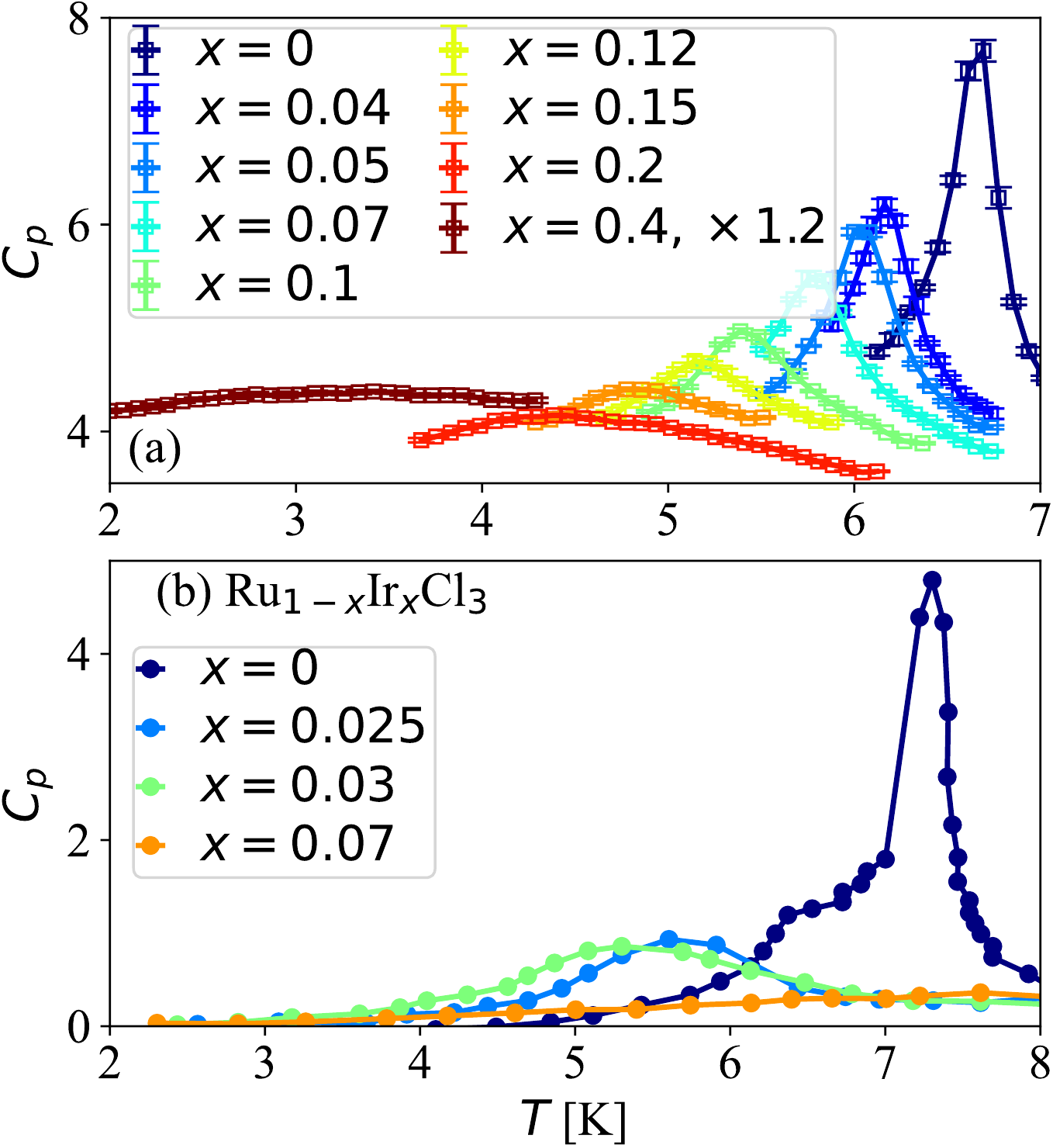}
\caption{\label{Fig:fig3} Magnetic specific heat $C_p$. Panel (a) shows simulated $C_p$ for various vacancy concentrations on a $2\times 64\times 64$ honeycomb lattice. Panel (b) plots the experimental measured specific heat of Ru$_{1-x}$Ir$_x$Cl$_3$, extracted from Ref.~\cite{LampenPRL2017}.}
\end{figure}

We take the spin configuration generated by Monte Carlo simulations on a $2\times 48\times 48$ lattice as an input of the Langevin equation. The fourth-order Runge-kutta method is used to evolve the spin configuration with a time step $\Delta t= 6.582$ fs. The initial $10^5$ time steps are used as thermalization, and the remaining $5\times 10^5$ time steps are set as a measurement window. Here, we are interested in the dynamical magnetic structure factor, which is obtained from
\begin{eqnarray}
S({\bf q},E)=\frac{1}{N_r} \sum_{\alpha\beta}\langle M_{\alpha}^r({\bf q},E) M_{\beta}^{r,*}({\bf q},E)\rangle,
\end{eqnarray}
where $M^r_\alpha({\bf q},E)=\sum_{\bf R}\int_0^{T} dt e^{i{\bf q}\cdot {\bf R}}e^{iE t} S_{{\bf R},\alpha}(t)$, $\alpha$ and $\beta$ denote the index of two sites in one unit cell, and $\langle \cdots \rangle$ denotes the average value of several time windows $[0, T]$. In our calculations, we set $T=3000\Delta t$.  Note that $r$ denotes the index of the random set. We use 108 different initial spin configurations with different defects. To capture quantum fluctuation, we follow Ref.~\cite{SamarakoonPRR2022} and renormalize $S(q,E)$ by $E\,n_B(E,T)$, where $n_B(E, T)$ is the bosonic distribution function.

\begin{figure}[t]
\center\includegraphics[width=0.95\columnwidth]{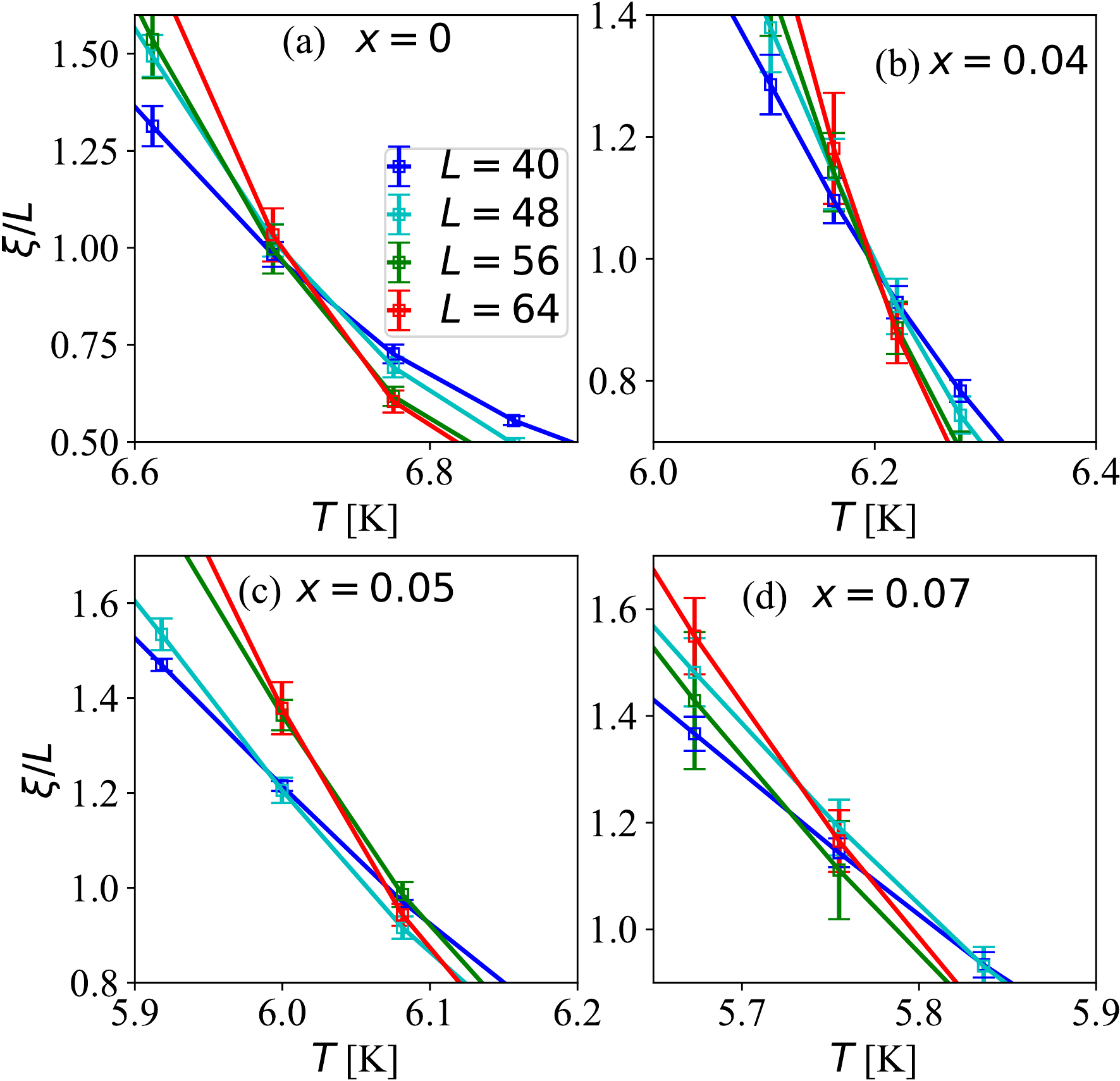}
\caption{\label{Fig:fig4} The finite-size scaling analysis of the zigzag spin correlation length $\xi$.}
\end{figure}

\section{Static magnetic properties}
We first focus on the static magnetic correlation function $\chi_m({\bf q})$ in the presence of vacancy defects at a low temperature $T=1$ K. Here, $\chi_m({\bf q})$ only includes correlations between the same sublattices. Fig.~\ref{Fig:fig1} plots $\chi_m({\bf q})$ for different vacancy concentrations with $L=48$. To provide a better visualization, $\chi_m({\bf q})$ is enhanced by factors of 2, 10, and 100 times in panels (b), (c), and (d), respectively. The ground state of our model with $x=0$ is the zigzag phase, leading to a strong peak appearing at the $M$ point in momentum space. Finite vacancy doping suppresses this zigzag state. The result shown in Fig.~\ref{Fig:fig1} is consistent with this prediction. Interestingly, we find that short-ranged zigzag correlations can persist up to a concentration that is larger than the site percolation threshold of a honeycomb lattice of $x=0.3$~\cite{FengPRE2008}, although the correlation strength is extremely weak. We also note that the real-space correlations in our MC simulations show that at $x=0.3$, the zigzag order breaks up into regions with three different directions of the zigzag chains (see appendix B).

\begin{figure*}[t]
\center\includegraphics[width=0.9\textwidth]{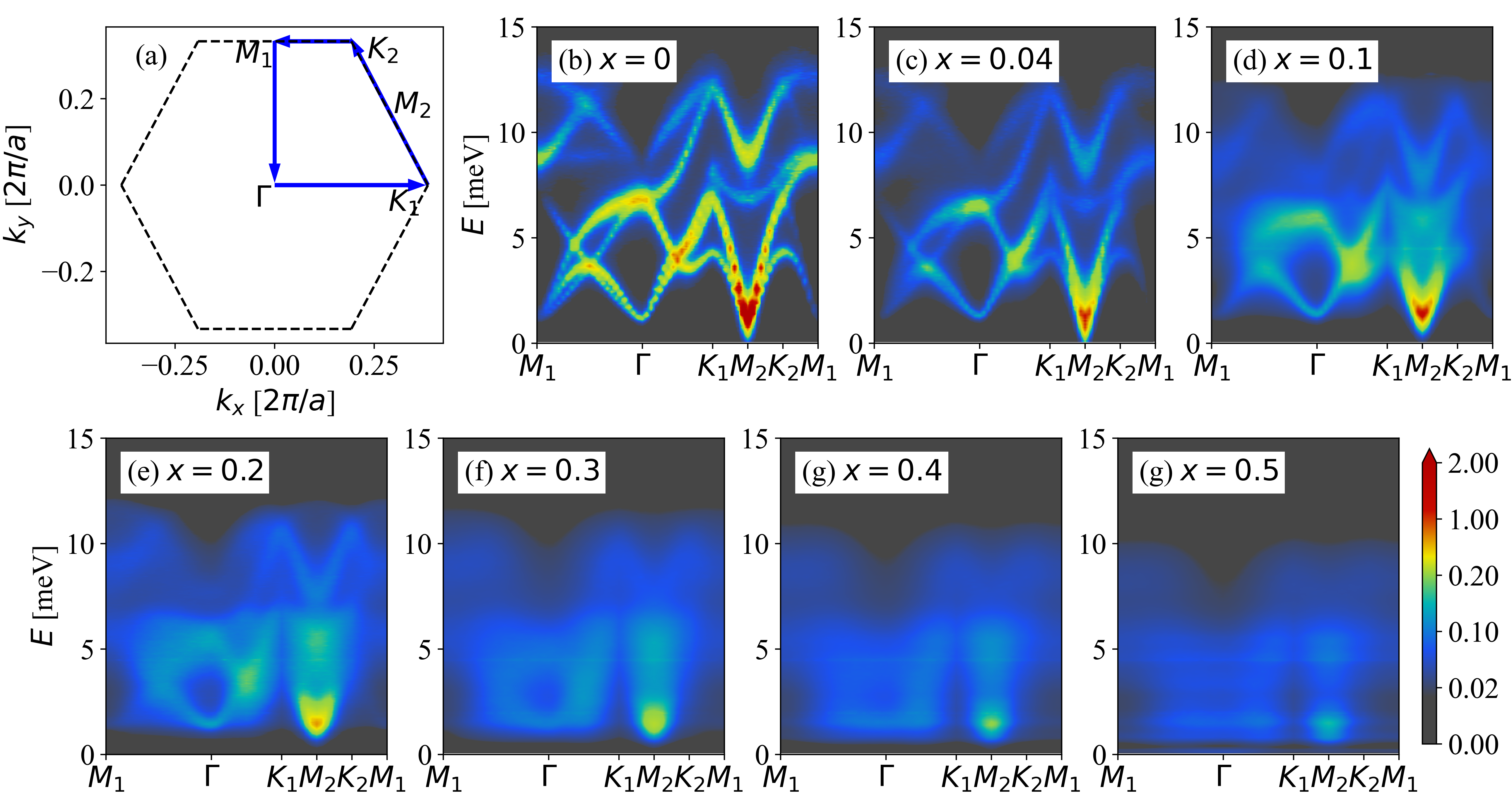}
\caption{\label{Fig:fig5} The evolution of the dynamical magnetic structure factor $S({\bf q},E)$ with different vacancy concentrations. Panel (a) sketches the high-symmetric path in the first Brillouin zone.}
\end{figure*}

An intriguing question regarding vacancy doping is the nature of the phase transition. Here, we plot the $x-T$ phase diagram in the low doping region in Fig.~\ref{Fig:fig2}, which is obtained by analyzing the specific heat $C_p$ and the correlation length $\xi$. The red diamond denotes the long-range ordering (LRO) temperature $T_N$ in the thermodynamic limit, obtained from the analysis of the scaling behavior described below.
The blue circle in Fig.~\ref{Fig:fig2} denotes the temperature $T_1$ where the specific heat has the maximum value. 
The suppression of the zigzag correlation is reflected by the doping-dependent behavior of $T_1$, which shows a linear decrease. We compare our results with experimental results on Ru$_{1-x}$Ir$_x$Cl$_3$, which exhibits two-phase transitions accompanied by a structural change to AB and ABC stackings. The temperatures for these phase transitions, extracted from Ref.~\cite{LampenPRL2017}, are plotted as cyan and green symbols in Fig.~\ref{Fig:fig2}.   
It is found that our theoretical result has a smaller slope. This inconsistency could be induced by three aspects, including quantum fluctuations, changes in the intralayer interaction induced by vacancies, and the spin-phonon or the electron-phonon interaction, which drives a structure change across the phase transition in Ru$_{1-x}$Ir$_x$Cl$_3$. We note that $T_N$ and $T_1$ are the same in the low doping region ($x<0.05$). In contrast, the long-range ordering temperature is absent when $x>0.05$.

To clarify the temperature-dependent behavior of $C_p$, we present detailed results on a $2\times 64\times64$ honeycomb lattice in Fig.~\ref{Fig:fig3}. When $x<0.05$, $C_p$ exhibits a $\lambda$-like shape, indicating the presence of a second-order phase transition. Further doping makes $C_p$ smooth around $T_1$, implying that $T_1$ cannot reflect a true phase transition. However, Fig.~\ref{Fig:fig1} shows significant zigzag spin correlations at $x=0.2$. Therefore, we infer that at $x>0.05$, $T_1$ denotes a transition temperature for the short-range order, which is labeled as SRO in Fig.~\ref{Fig:fig2}. The evidence of the absence of the long-range order will be discussed later. 
Experimentally, a similar doping-dependent behavior of $C_p$ is observed in Ru$_{1-x}$Ir$_x$Cl$_3$~\cite{DoPRB2018}. However, we must clarify that the long-ranged order defined in experiments is based on the anomalous behavior (the dome structure) of the specific heat and uniform magnetic susceptibility. Our numerical results show that this definition does not correctly reflect the long-range order physics because the anomalous behavior also exists in the short-range ordered state.

When the vacancy concentration exceeds the honeycomb lattice site percolation threshold ($x>0.3$), the dome structure of $C_p$ becomes very flat, making it difficult to find $T_1$. Consequently, Fig.~\ref{Fig:fig2} only shows results for $x\le0.3$. Fig.~\ref{Fig:fig3}(b) shows the specific heat of the second phase transition in Ru$_{1-x}$Ir$_x$Cl$_3$, which has a lower transition temperature. These results are extracted from Ref.~\cite{LampenPRL2017}. Compared to the first phase transition, the second phase has a sharper signature at $x=0$, similar to our theoretical result. Fig.~\ref{Fig:fig3} (b) shows that $C_p$ loses the $\lambda$-shape feature at $x=0.025$. Therefore, a tiny vacancy doping can destroy the long-range ordered state in the Ru$_{1-x}$Ir$_x$Cl$_3$ sample. Compared with our theoretical results, the experimentally observed long-range order is more fragile.

We use the finite-size scaling theory to precisely determine the temperature of the phase transition in the thermodynamic limit~\cite{KimPRE1996,JonesPRB2005}. In the thermodynamic limit, $\xi/L$~\cite{BallesterosPRB2000} is independent of the lattice size at the critical temperature. Fig.~\ref{Fig:fig4} shows the temperature-dependent correlation length $\xi/L$ for four different doping concentrations. At $x=0$ and $x=0.04$, $\xi/L$ crosses a single point for four different lattice sizes with an error smaller than 0.01 K. Here, we define the temperature at the crossing point as the long-range ordering temperature $T_N$. At $x=0.05$ ($x=0.07$), the curves for $L=64$ and for $56$ cross around $T=6$ K (5.85 K), and the curves for $L=64$ and $48$ cross around $T=6.1$ K (5.75 K). Compared to the small error in the $x=0$ and $0.04$ cases, we deduce that there is no long-range order for $x\ge0.05$ in the thermodynamic limit~\cite{VietPRL2009}. Since the scaling behavior for $x=0.04$ and $0.05$ are significantly different, we believe that the absence of a single crossing point in Fig~\ref{Fig:fig4} (c) is not due to a numerical instability. In fact, the same analysis and conclusion have been made in the previous study of the pure Kitaev-Heisenberg model and the $J_1$-$J_2$-$J_3$ model~\cite{AndradePRB2014}. 


\begin{figure}[t]
\center\includegraphics[width=\columnwidth]{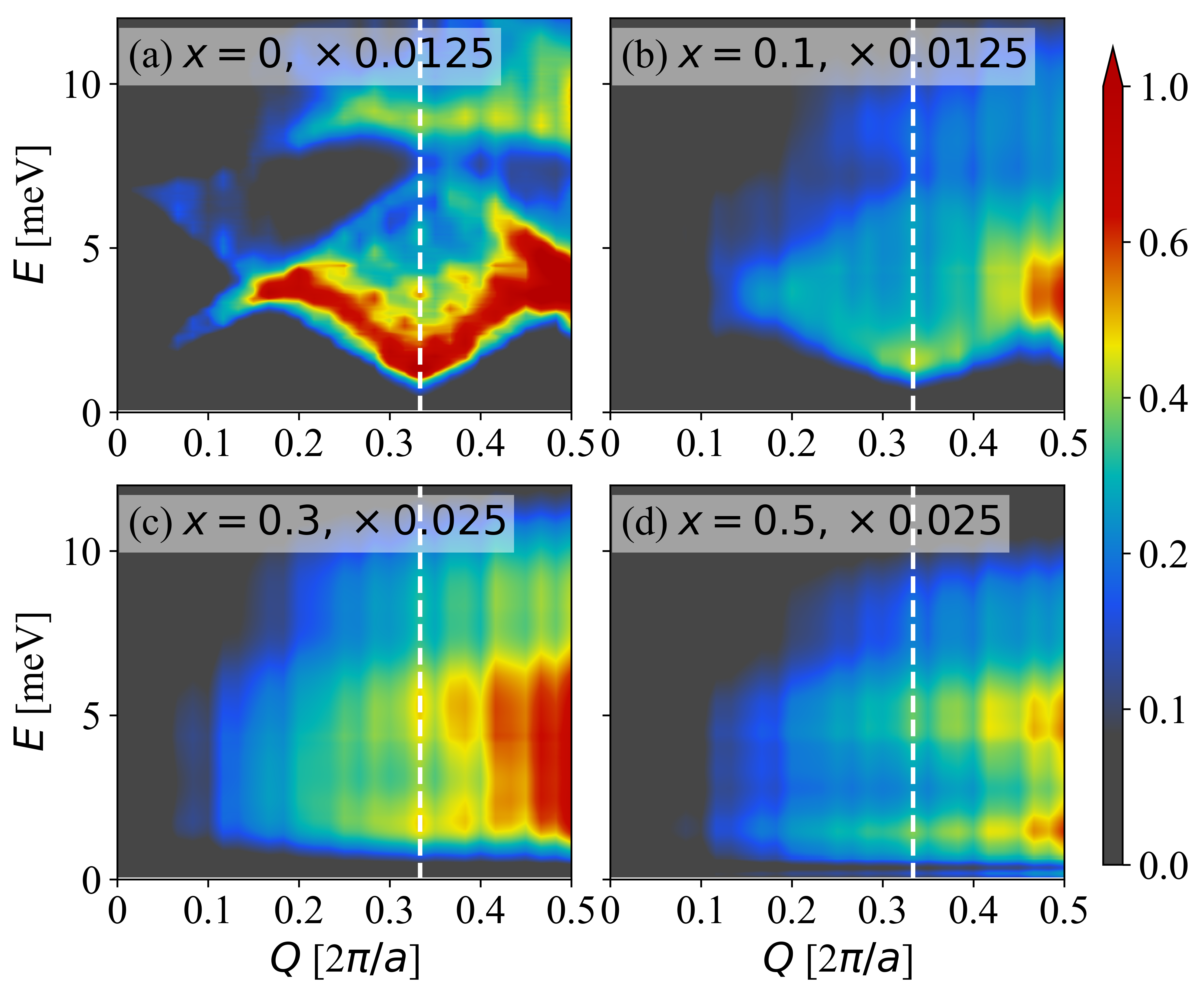}
\caption{\label{Fig:fig6} The dynamical magnetic structure factor $I(Q,E)$ for the polycrystal. The white dashed line shows the momentum length of the $M$ point. The intensity in panels (a), (b), (c), and (d) is scaled by 0.0125,0.0125, 0.025, and 0.025, respectively.}
\end{figure}

\section{dynamical magnetic properties}
Fig.~\ref{Fig:fig5} shows the dynamical magnetic structure factor $S(q,E)$ along the high-symmetric path at $T=1$ K, which is sketched in panel (a). At $x=0$, we use 108 sets of input spin configurations with different wave vectors. In the static spin correlation function for a single sublattice $\chi_m(q)$, plotted in Fig.~\ref{Fig:fig1}, we obtained six Bragg peaks located at the $M$ points. However, the spin correlation between the same sublattice and different sublattices on the honeycomb lattice have opposite signs for the zigzag state, causing the total spin correlation function to vanish at the $M_1$ points. Therefore, we only observe the low-energy magnon mode at the $M_2$ point in panel (b). The corresponding spin gap of these modes is 1.3 meV. 

As the vacancy concentration increases, the sharp signature of the dynamical magnetic structure factor becomes smooth and broadens. Although the magnetic structure intensity is significantly suppressed by vacancy doping, the magnon mode at the $M_2$ point is robust due to the locally ordered spins. In fact, the low-energy magnon mode persist all the way to $x=0.5$, far beyond the site percolation threshold at $x=0.3$. In addition, we also observe continuous spin excitations induced by doping, which first appear around the $M_2$ point and then arise around the $\Gamma$ point. This continuous spin excitation reflects the disorder scattering of spins to high energy states at each momentum.
In addition, the excitation energy for the low-energy mode at the $M_2$ point decreases with doping when $x<0.04$ and increases with further doping (see appendix B). This behavior is consistent with experimental observations in Ru$_{1-x}$Ir$_x$Cl$_3$~\cite{LampenPRL2017}. Around the $\Gamma$ point, the spin excitation energies are softened and become independent of momentum at large vacancy concentrations ($x>0.3$).

\begin{figure}[t]
\center\includegraphics[width=0.95\columnwidth]{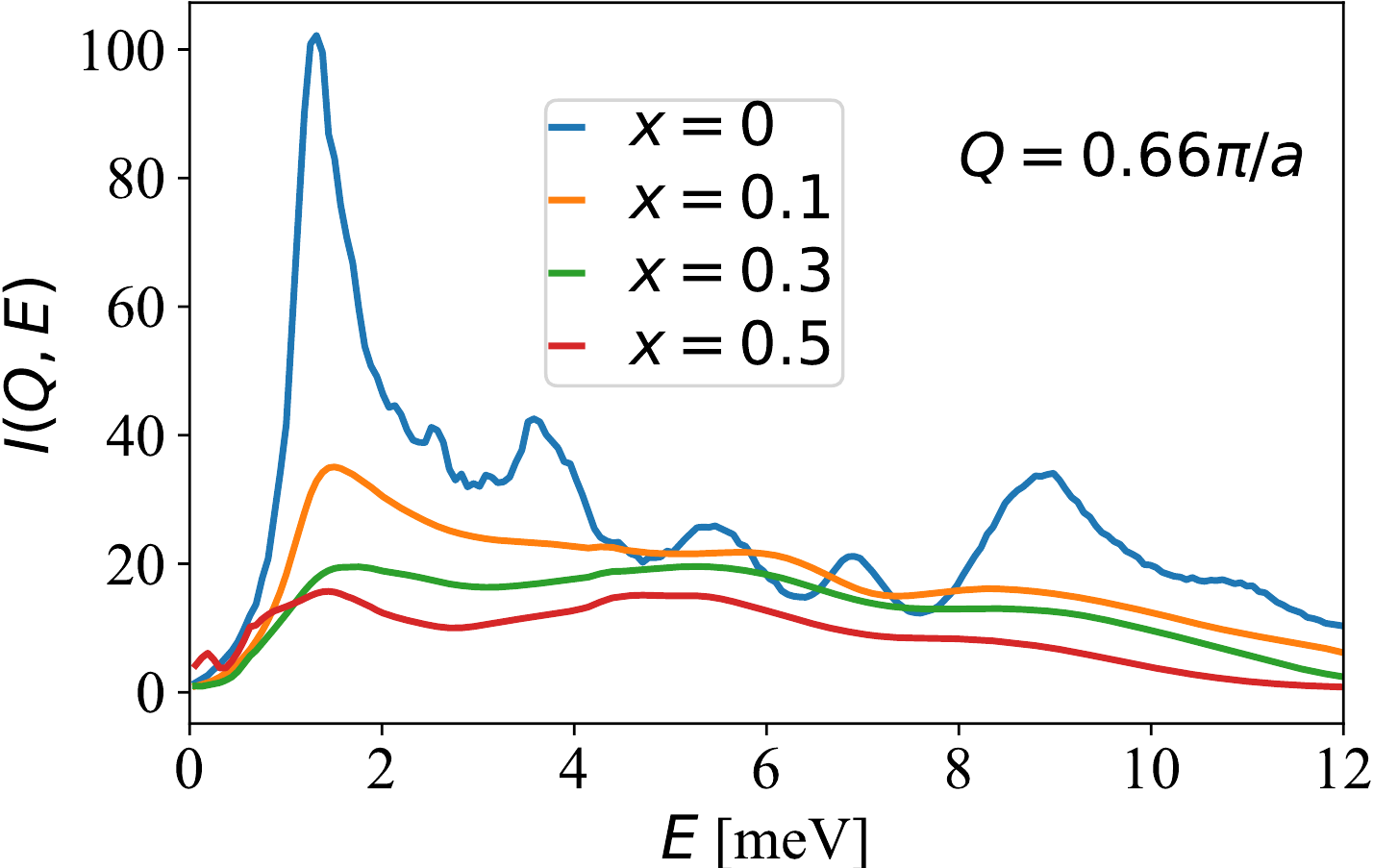}
\caption{\label{Fig:fig7} The dynamical magnetic structure factor $I(Q,E)$ in a polycrystal at $Q=0.66\pi/a$.}
\end{figure}

A previous inelastic neutron scattering experiment on the polycrystal Ru$_{1-x}$Ir$_x$Cl$_3$ reported two modes in the low energy region around momentum amplitude $Q=0.66\pi/a$. One mode at 4 meV was attributed to the magnons and the other at 6 meV to fractional excitations. While the feature associated with magnons vanishes at $x=0.35$, the feature associated with fractional excitations persist
at $x=0.35$. 

To compare with experimental results, we plot the dynamical magnetic structure factor $I(Q,E)$ for the polycrystal in Fig.~\ref{Fig:fig6}, which is obtained by summing $S({\bf q},E)$ over all ${\bf q}$ with $|{\bf q}|=Q$. The white dashed line in Fig.~\ref{Fig:fig6} shows the momentum amplitude of the $M$ point. For better visualization, we plot $I(Q=0.66\pi/a,E)$ in Fig.~\ref{Fig:fig7}. At $x=0$, we observe many modes in the low energy region due to the sharp signature generated by our theoretical simulations. The lowest-energy magnon mode is located at 1.3 meV, higher than the experimental result. This inconsistency arises because our model was derived for the single crystal in Ref. ~\cite{SamarakoonPRR2022}, instead of the polycrystal in Ref.~\cite{LampenPRL2017}. As the vacancy concentration increases, the 1.3 meV magnon mode is suppressed but does not vanish. When $x\ge0.3$, the magnon modes at 1.3 meV and 5 meV have a similar intensity. 

\section{Conclusion}
In summary, we study the magnetic properties of the extended Kitaev-Heisenberg model with spin vacancies and find that the local zigzag correlation can persist up a concentration that is larger than the percolation threshold. Both static and dynamic results support this conclusion. We analyze the specific heat and the correlation length and find that the long-range zigzag order vanishes as the doping concentration increases to 5\%. Meanwhile, the ground state exhibits short-range order. We also examine the dynamical magnetic structure factor, which shows that the low-energy magnon mode at the $M_2$ point persists in the short-range ordered state, although its intensity is significantly suppressed. Our results can help interpret existing inelastic neutron experiments on polycrystalline Ru$_{1-x}$Ir$_x$Cl$_3$ samples, and future inelastic neutron experiments on Ru$_{1-x}$Ir$_x$Cl$_3$ single crystals and diluted Kitaev candidate materials in general. Besides, we observe two inconsistencies between our theoretical predictions and experimental results. First, our $T_1$ decreases faster than the experimental results. Second, our local zigzag correlation is more robust than that of the experimental correlation. These inconsistencies can be attributed to the absence of quantum fluctuations, interlayer interactions, spin-phonon couplings and modifications of exchanges between non-vacant sites in our simulations. 


\section{acknowledgments}
This research was supported by the U.S. Department of Energy, Office of Science, National Quantum Information Science Research Centers, Quantum Science Center. 
This research used resources of the Compute and Data Environment for Science (CADES) at the Oak Ridge National Laboratory, which is supported by the Office of Science of the U.S. Department of Energy under Contract No. DE-AC05-00OR22725. 

\section{Appendix A: Real space spin correlation}
\begin{figure*}[t]
\center\includegraphics[width=0.9\textwidth]{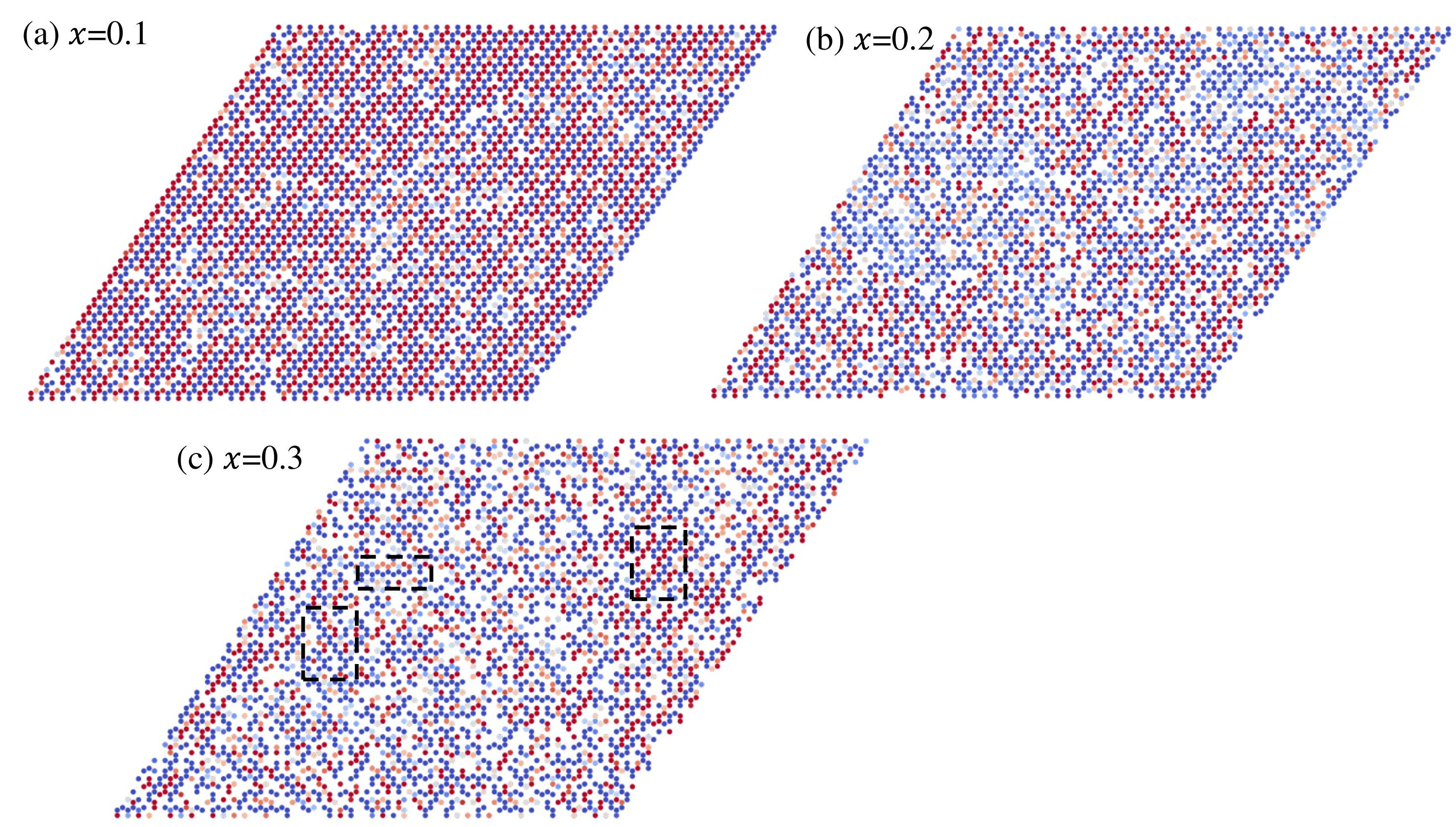}
\caption{\label{Fig:fig8} Snapshot real space spin correlations for $x=0.1$, $x=0.2$, and $x=0.3$ at $T=1$. Here blue,white and red interpolate between $\chi({\bf r}, {\bf r_0})=1$, $\chi({\bf r}, {\bf r_0})=0$, and $\chi({\bf r}, {\bf r_0})=-1$, respectively. }
\end{figure*}

Here, we present snapshot real space spin correlations $\chi({\bf r}, {\bf r_0})$ from MC simulations. $\chi({\bf r}, {\bf r_0})$ is defined as 
\begin{eqnarray}
\chi({\bf r}, {\bf r_0})=\bf{S}_{\bf r} \cdot \bf{S}_{{\bf r}_0},
\end{eqnarray}
where ${\bf r}$ is the position of the spin site and ${\bf r}_0$ is the position of the reference site.
Fig.~\ref{Fig:fig8} shows results for $x=0.1$, $x=0.2$, and $x=0.3$ at $T=1$ K. Here, we set ${\bf r}_0=0$. The red color denotes the positive correlation, and the blue color denotes the negative correlation. At $x=0.1$, all spins are aligned along one wave vector. At $x=0.3$, close to the site percolation treshold of the honeycomb lattice, the local spins can align along different wave vectors in different regions. We highlight these local spin correlations with dashed black rectangles in Fig.~\ref{Fig:fig8}.

\section{Appendix B: dynamical magnetic structure at $Q=M_2$}
\begin{figure}[t]
\center\includegraphics[width=0.9\columnwidth]{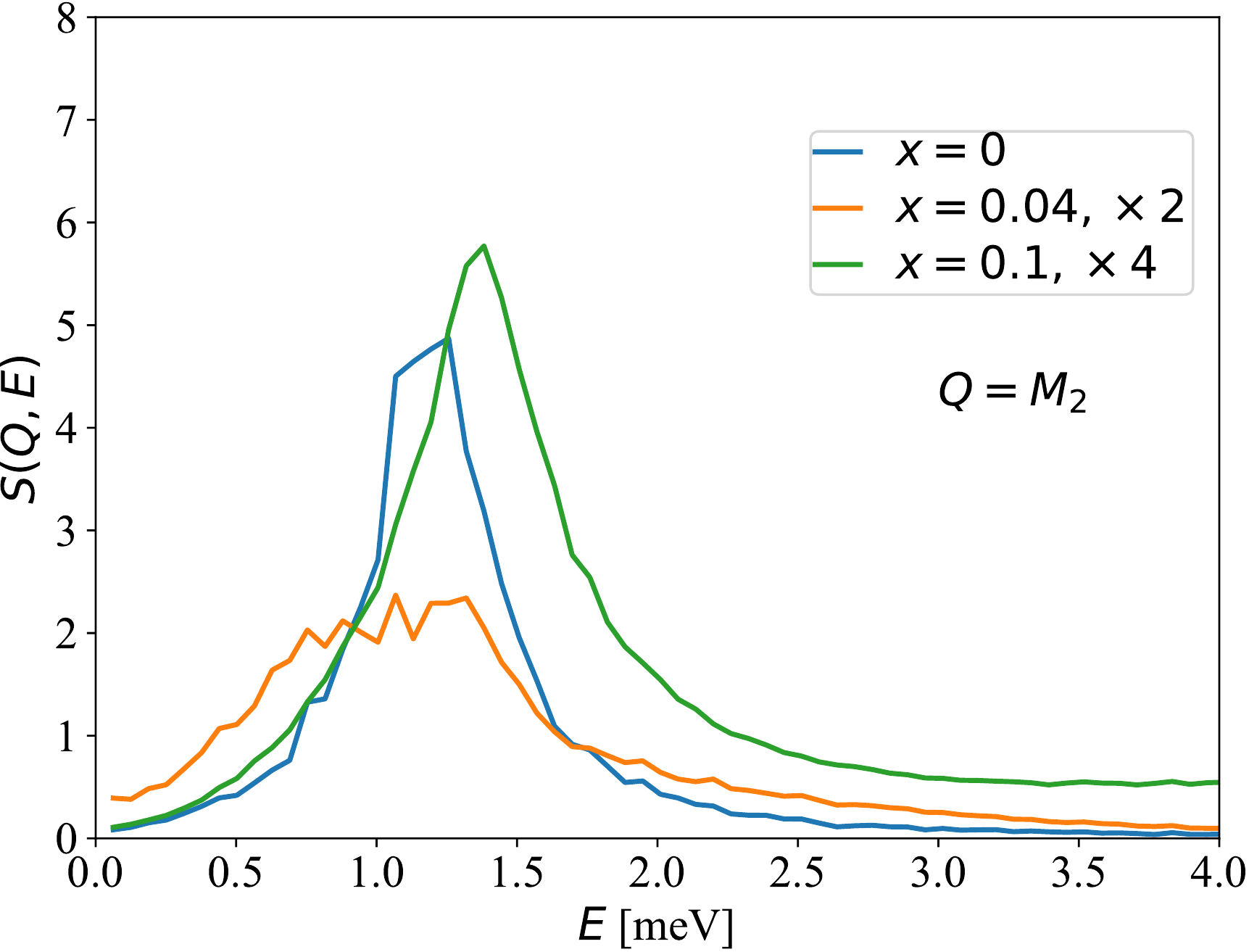}
\caption{\label{Fig:fig9} Dynamical magnetic structures at $Q=M_2$ for $x=0$, 0.04, and 0.1, respectively.}
\end{figure}

Fig.~\ref{Fig:fig9} shows the dynamical magnetic structures at $Q=M_2$ for $x=0$, 0.04, and 0.1, respectively. 
It is found that a small doping ($x=0.04$) broadens the signature of the magnetic structure and lowers the spin excitation energy. By further increasing doping, the magnetic structure exhibits a sharp peak, and the spin excitation energy increases.

\bibliography{main}

\end{document}